\documentclass[aps,prl,twocolumn,showpacs,groupedaddress]{revtex4}
\usepackage{amsmath}
\usepackage[dvips]{graphicx} 
\usepackage{amssymb}
\usepackage{color}
\usepackage{float}
\usepackage{amsmath,amssymb,graphicx,epsfig}
\usepackage[english]{babel}
\usepackage{epstopdf}
\makeatletter
\newcommand*{\balancecolsandclearpage}{%
  \close@column@grid
  \cleardoublepage
  \twocolumngrid
}
\makeatother

\newcommand{\Tau}{\mathrm{T}}
\newcommand{\nobracket}{}

\newcommand{\tmop}[1]{\ensuremath{\operatorname{#1}}}

\newcommand{\tmrsup}[1]{\textsuperscript{#1}}

\begin{document}

\title{Fluctuation-dissipation relations of a tunnel junction driven by a quantum circuit}

\author{O. \surname{Parlavecchio}$^{1}$}

\author{C. \surname{Altimiras}$^{1}$}
\email{email: carles.altimiras@sns.it}
\altaffiliation[Present address: ]{NEST, Istituto Nanoscienze CNR and Scuola Normale Superiore, Piazza San Silvestro 12, 56127, Pisa, Italy}

\author{J.-R. \surname{Souquet}$^{2}$}
\altaffiliation[Present address: ]{Department of Physics, McGill University, Montr´eal, QC, Canada}

\author{P. \surname{Simon}$^{2}$}

\author{I. \surname{Safi}$^{2}$}              

\author{P. \surname{Joyez}$^{1}$}

\author{D. \surname{Vion}$^{1}$}

\author{P. \surname{Roche}$^{1}$}
 
\author{D. \surname{Est\`eve}$^{1}$}

\author{F. \surname{Portier}$^{1}$}
\email{email: fabien.portier@cea.fr}

\affiliation{$^{1}$ SPEC (UMR 3680 CEA-CNRS), CEA Saclay, 91191 Gif-sur-Yvette, 
France}
\affiliation{$^{2}$ Laboratoire de Physique des Solides, Universit\'e Paris-Sud, 91405 Orsay, France}

\begin{abstract}
We derive fluctuation-dissipation relations for a tunnel junction driven through a high impedance microwave resonator, displaying strong quantum fluctuations. We find that the fluctuation-dissipation relations derived for classical forces hold, provided the effect of the circuit's quantum fluctuations is incorporated into a modified non-linear current voltage chaacteristics.  We also demonstrate that all quantities measured under a time dependent bias can be reconstructed from their values measured under a dc bias using photo-assisted tunneling relations. We confirm these predictions by implementing the circuit and measuring the dc current through the junction, its high frequency admittance and its current noise at the frequency of the resonator.
	
\end{abstract}

\date{\today}


\maketitle

The fluctuations of any physical system held at equilibrium are proportional to its dissipative linear response\cite{JohnsonNoise,NyquistNoise,CallenWelton,Kubo}. This universal fluctuation-dissipation theorem (FDT) relates the fluctuations (noise) of any passive system to an easier linear response measurement. Moreover, measuring both quantities implements a primary thermometer, reaching metrological accuracy \cite{0026-1394-48-3-008}. The FDT provides a physical picture on the origin of macroscopic irreversibility: it shows that dissipation within a Hamiltonian system corresponds to the system's ability to dissolve an incoming excitation within its internal degrees of freedom through the system's fluctuations of both quantum and thermal origin. Turned into a rigorous formalism as in \cite{Caldeira-Leggett}, such a picture provides an efficient way to model dissipation in macroscopic quantum systems. Despite these successes, the standard FDT \cite{CallenWelton} relates the fluctuations of a quantum system to its response to a classical drive, ignoring the quantum fluctuations of the driving forces. In the particular case of a quantum electrical conductor connected to an electromagnetic environment, these quantum fluctuations are known to trigger inelastic electron tunneling, yielding a nonlinear dc current-voltage characteristics $I(V_{dc})$, an effect known as Environmental or Dynamical Coulomb Blockade (DCB) \cite{ingoldnazarov1992DCB}. This raises the question of the existence of FDT relations for a quantum conductor coupled to an environment displaying strong quantum fluctuations. Such relations have been derived \cite{Safi-Joyez, SafiTunnel,LeeLevitovDCBPRB1996}, but the bias was described as a time-dependent classical voltage across the junction. Here instead, we explicitly include the quantum fluctuations of the voltage across a normal tunnel junction biased through the elementary building block of the description of a linear circuit, i.e. an harmonic oscillator, driven in a coherent state.  We find that in this case, the junction's response and current fluctuations can be recast in terms of the nonlinear $I(V_{dc})$ curve. We thus extend the validity of expressions found for tunnel conductors driven by classical fields \cite{SafiTunnel, LeeLevitovDCBPRB1996, SafiSukho, Safi-Joyez, DahmJosephsonLinewidthPRL1964, RogovinScalapino1974,SukhoLossCotunnNoisePRB2001,TienGordonPR1963,PhysRevLett.72.538,PhysRevB.58.12993}.  We probe these predictions by embedding a tunnel junction in a high impedance microwave resonator. The junction's dc conductance, its finite frequency admittance and its current fluctuations are found in good agreement with predictions.

\begin{figure}[t]
  \includegraphics[width=8cm]{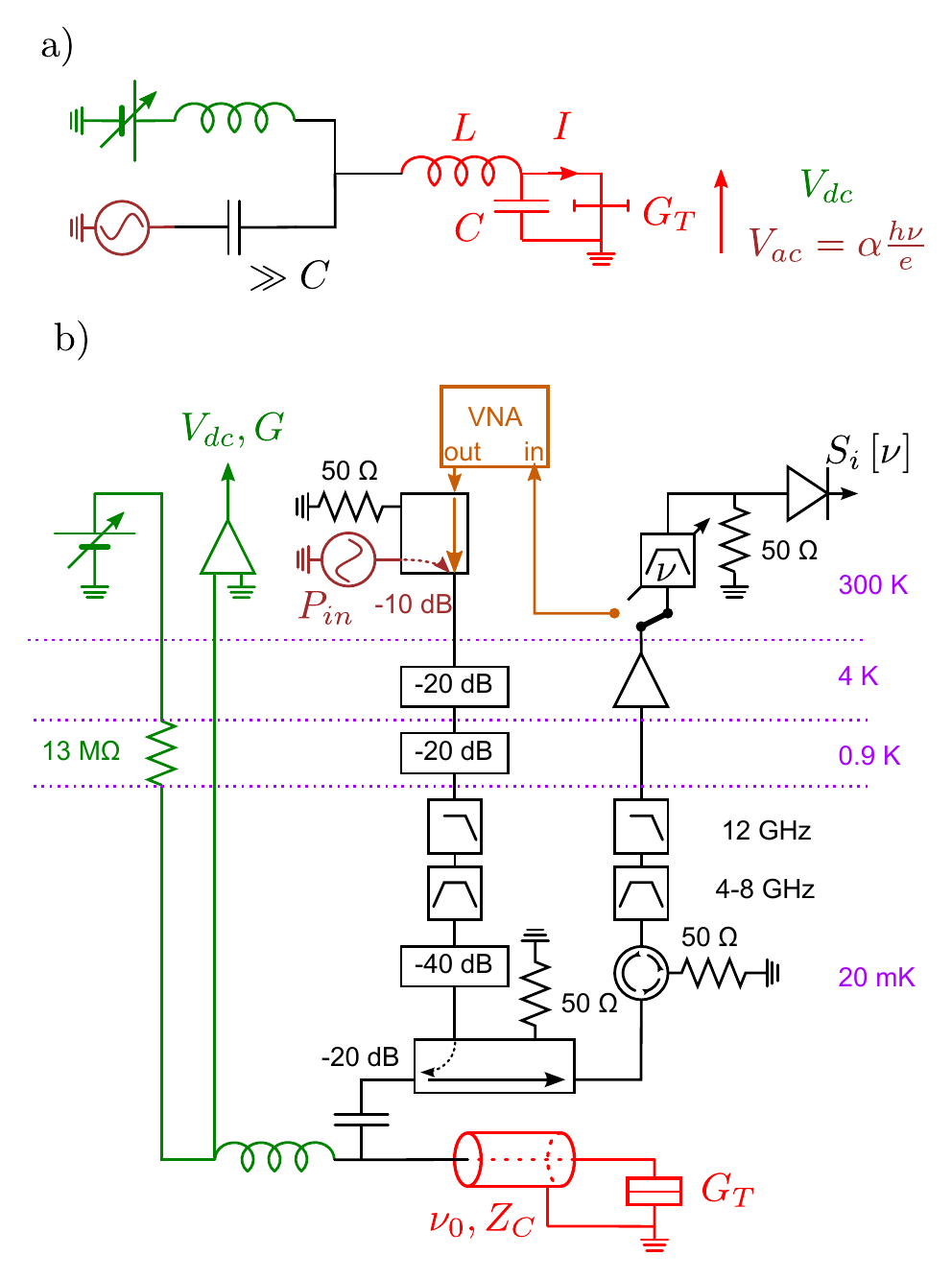}
  \caption{ (color online) a) Model system: a small conductance tunnel junction $G_T$ is embedded in a resonator of frequency $\nu_0=1/\sqrt{L C}$ and characteristic impedance $Z_C=\sqrt{L/C}$. The system is connected to a dc (rf) voltage source $V_{dc}$ ($V_{ac}$) through a large inductance (capacitance). b) Experimental set-up: A normal tunnel junction, cooled at 20 mK by a dilution refrigerator, is connected to a 50~$\Omega$-line through a high-impedance $\lambda/4$-resonator, whose inner conductor consists in a serial SQUID array. The resonator is connected to a bias-Tee, whose inductive port allows us to dc bias the junction and to measure its low frequency conductance. The RF port allows us to shine microwaves onto the resonator, and to measure the microwave signals emitted/reflected by the sample.}
	\label{Figure1}
\end{figure}

We consider (see Fig. \ref{Figure1}-a) a tunnel junction of tunnel conductance $G_T$  embedded at temperature $T$ in an $LC$ circuit with resonant frequency $\nu_0=1/( 2 \pi\sqrt{LC})$ and characteristic impedance $Z_C=\sqrt{L/C}$, where $C$ is the oscillator's capacitance, and $L$ its inductance. Harmonic oscillators have been considered as detectors for the current fluctuations of quantum conductors \cite{LoosenLesovikJETP1997,PhysRevB.62.R10637,PhysRevLett.99.066601, PhysRevB.81.205411}, but neglecting the back-action due to their quantum voltage fluctuations. We assume that the quantum average of the voltage across the junction reads $V_{dc} + V_{ac}\cos(2\pi \nu_0t)$. We describe the resonator field by a 'thermal coherent state' density matrix $\rho=D(i\alpha/2r)\rho_T D(i\alpha/2r)^\dagger$\cite{doi:10.1080/09500348814550571}, where $\rho_T$ is the equilibrium density matrix at temperature $T$, $D(i\alpha/2r)=\exp\left[i\alpha(a+a^\dagger)/2r\right]$ is the displacement operator corresponding to an amplitude $i\alpha/2r$, with $\alpha=eV_{ac}/h\nu_0$ and $r=\sqrt{\frac{\pi Z_c}{R_K}}$ ($R_K=h/e^2$ = 25.8 k$\Omega$) characterizing the coupling between the oscillator and the tunnel junction. We evaluate \cite{SM} the time-dependent quantum average of the current $I$ and current fluctuations spectral density $S_I$ to lowest order in the tunnel coupling. Their time average $\overline I$ and  $\overline {S_I}$ are obtained as copies of the same quantity measured under dc bias, translated by the different harmonics of $\nu_0$ and weighted by Bessel functions, following photo-assisted tunneling relations \cite{TienGordonPR1963,PhysRevLett.72.538,PhysRevB.58.12993,SafiSukho, SafiTunnel}:

\begin{eqnarray}
\overline I(V_{dc},\alpha)=\sum_k J_k(\alpha)^2 I(V_{dc}-kh\nu_0/e, 0) \label{eq:I-Tien-Gordon}\\
\overline {S_I}(\nu,V_{dc},\alpha)=\sum_k J_k(\alpha)^2 S_I(\nu,V_{dc}-kh\nu_0/e,0). \label{eq:SI-Tien-Gordon}
\end{eqnarray}

The time dependence of $I$ and $S_I$ can also be retained to calculate their Fourier transform, allowing the derivation of
the junction's admittance $Y(\nu,V_{dc},\alpha)$ from the current response to an infinitesimal drive in a small impedance additional fictitious mode at an arbitrary frequency $\nu$ \cite{SM}. We find that $\operatorname{Re}[Y(\nu,V_{dc},\alpha)]$ obeys a photo-assisted tunneling formula analog to Eqs. \ref{eq:I-Tien-Gordon}-\ref{eq:SI-Tien-Gordon} and keeps a structure known for classical drives \cite{RogovinScalapino1974,SukhoLossCotunnNoisePRB2001,SafiSukho,SafiTunnel}:

\begin{equation}
\operatorname{Re}[\overline{Y}(\nu,V_{dc},\alpha)]=e\frac{\overline I(V_{dc}+h\nu/e,\alpha)-\overline I(V_{dc}-h\nu/e,\alpha)}{2h\nu}
\label{Ydenu}
\end{equation}
whereas its imaginary part follows from Kramers-Kronig relations. From a similar calculation for the noise spectral density we recover the noise susceptibility derived in \cite{GabelliNoiseSensiPRL2008} using the Landauer-B\" uttiker formalism. 

For a dc bias $\alpha=0$, the current \cite{ingoldnazarov1992DCB} and current noise \citep{AltimirasNoiseDCBPRL2014} read: 
\begin{eqnarray*}
  I (V_{dc}) & = & \frac{G_{\text{T}}}{e}  [ \nobracket \gamma \ast P(eV_{dc})- \gamma
  \ast P( -eV_{dc})], \label{EqIDCB} \\                                                                   
  S_{I} ( \nu ,V_{dc}) & = & 2G_{\text{T}}  [ \gamma \ast P(eV_{dc}-h \nu )+ \gamma
  \ast P( -h \nu -eV_{dc})] \label{EqSIDCB}, 
\end{eqnarray*}
where $\gamma \ast P (E) = \int d \varepsilon'
\gamma ( \varepsilon' ) P (E- \epsilon' )$ with $P ( \varepsilon )$ the
probability density for a tunneling electron to emit the energy $\varepsilon$
in form of photons into the impedance {\cite{ingoldnazarov1992DCB}}, with
$\gamma ( \epsilon ) = \int d \varepsilon' f ( \varepsilon' )  [1-f(
\varepsilon' + \varepsilon )] = \varepsilon / (1-e^{- \varepsilon /k_{B} T}
)$, and $f$ the Fermi function. Combining these expressions with Eqs.\ref{eq:I-Tien-Gordon}-\ref{Ydenu}, we obtain a Kubo-like relation \cite{Kubo,Safi-Joyez}:  

\begin{equation*}
\overline{S_I}(-\nu, V_{dc},\alpha)-\overline{S_I}(\nu, V_{dc},\alpha)=2h\nu \operatorname{Re} \overline Y(\nu,V_{dc},\alpha).
\end{equation*}

The detailed balance property of $\gamma(E)$ and $P(E)$ yields:

\begin{equation}
\overline{S_I}(\nu,V_{dc},\alpha)=\frac{e\overline I\left(V_{dc}-h\nu,\alpha\right)}{1-e^{-\beta\left(eV_{dc}-h\nu\right)}}+\frac{e\overline I\left(V_{dc}+h\nu,\alpha\right)}{e^{-\beta\left(-eV_{dc}-h\nu\right)}-1}. \label{FDT_I-EmissionNoise}
\end{equation}

Eq. (\ref{FDT_I-EmissionNoise}) is the main prediction we probe experimentally: The results derived for classically dc biased tunnel elements \cite{RogovinScalapino1974,PhysRevLett.96.136804,SukhoLossCotunnNoisePRB2001} can be extended to quantum biasing circuits, provided one incorporates the effect of their quantum fluctuations into a "renormalized" non-linear $I(V_{dc})$ curve \cite{LeeLevitovDCBPRB1996}, even in the presence of a time-dependent drive \cite{SafiTunnel} since it modifies equally both current and noise. The crucial assumptions of our derivation are i) that the quantum conductor is in the tunnel regime with  ii) a tunnel conductance small enough to have negligible impact on the density matrix of the system, which iii) follows a detailed balance. Note that in references \cite{DahmJosephsonLinewidthPRL1964, RogovinScalapino1974,SukhoLossCotunnNoisePRB2001,LeeLevitovDCBPRB1996}, only the symmetrized spectral density of the current fluctuations $\left[S_I(\nu, V_{dc})+S_I(-\nu, V_{dc})\right]/2$ was considered having a similar yet different expression.

\begin{figure*}[p!]
\centering
\includegraphics[width=1.0\textwidth]{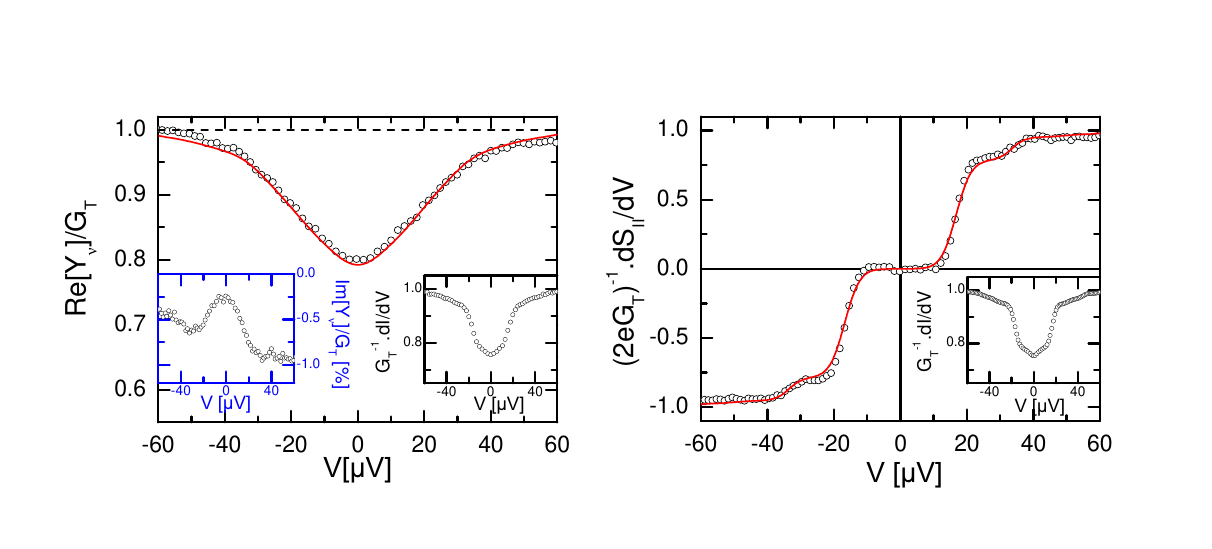}
  \caption{ (color online) 
 \textit{Left panel}: (Black circles) Admittance computed from reflection measurement at the frequency of the resonator $\nu_0=4.1\,\mathrm{GHz}$ for a small RF drive ($\alpha \ll 1$), as a function of the dc-voltage bias.  \textit{Right panel}: (Black circles) Derivative of the noise with respect to dc bias, measured at resonant frequency, $\nu_0=4.0\,\mathrm{GHz}$, as a function of dc-voltage bias. 
(Red curves) Theoretical curves computed from Eq.~(\ref{Ydenu}) and Eq.~(\ref{FDT_I-EmissionNoise}), using the (black right insets) 
 dc-conductance, which shows step-like features characteristic of the DCB by a single mode. (Left panel, left inset) Estimated susceptance of the tunnel junction.}
\label{Figure2} 
\end{figure*}

\begin{figure*}[p!]
\centering
  \includegraphics[width=6.6in]{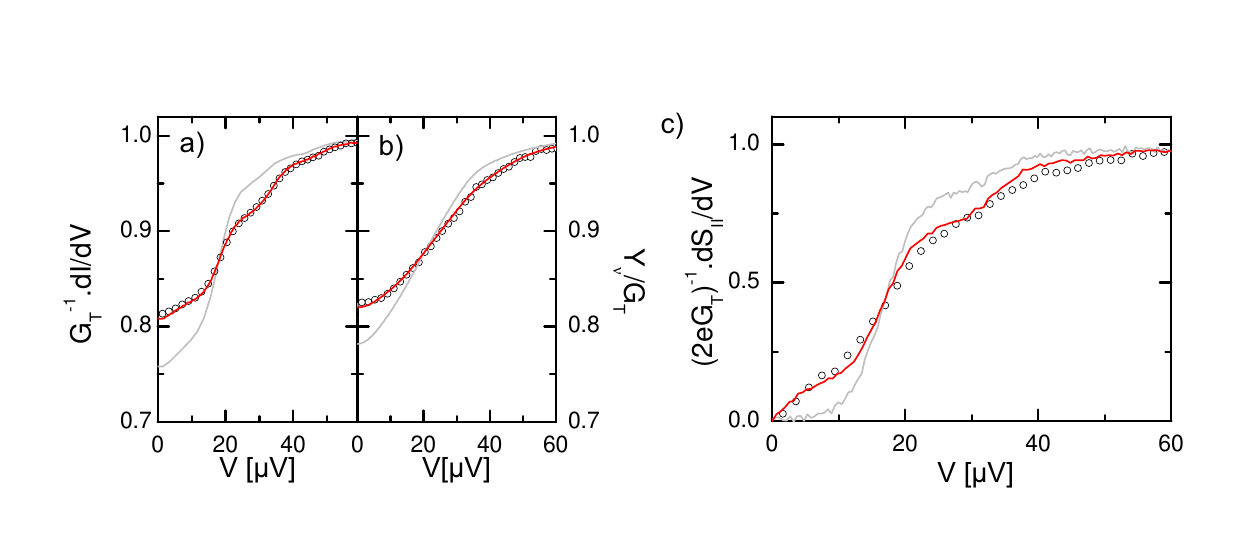}
  \caption{(color online) Photo-assisted regime: Black circles show  (a)
the junction's conductance at low frequency, (b)  at the resonator's frequency $\nu_0=4.1$ GHz, and  (c) the derivative of the emission current noise with respect to the bias voltage, measured in the presence of an RF drive ($\alpha=1.15$). Red curves show the photo-assisted tunneling predictions Eqs.(\ref{eq:I-Tien-Gordon}-\ref{eq:SI-Tien-Gordon}), using the same quantity measured for $\alpha=0$ (shown by the grey lines).} 
\label{Figure3} 	
\end{figure*}

To probe the above predictions, we implement a circuit equivalent to Fig. \ref{Figure1}-a) by embedding a high impedance tunnel junction in a microwave resonator with an impedance high enough to  significantly modify the tunnel junction transport properties \cite{AltimirasHighZMatchAPL2013, AltimirasNoiseDCBPRL2014}. We then measure both the dc conductance, the finite frequency admittance and the finite frequency current noise of the junction as a function of its dc voltage bias and in the presence of an additional microwave tone inducing a coherent state in the resonator, using a cryogenic low noise amplifier. Our setup only gives access to the emission noise spectral density of the sample \cite{ AltimirasNoiseDCBPRL2014}, unlike on-chip quantum detectors \cite{AguadoQDetPRL2000,PhysRevLett.96.176601,PhysRevLett.99.206804} such as SIS detectors \cite{Deblock11072003} which allow the measurement of the emission and absorption noise of their linear environment \cite{PhysRevLett.105.166801}. Coupling two such detectors via a low impedance circuits, displaying negligible quantum fluctuations, was used to measure their emission noise \cite{PhysRevLett.105.166801}, found in agreement with Eq.\ref{FDT_I-EmissionNoise}, and to probe their absorption noise \cite{PhysRevLett.96.136804} and admittance \cite{PhysRevB.85.085435}.

Our experimental set-up is represented in Fig.\ref{Figure1}-b): a 100 $\times$ 100 nm$^2$ tunnel
junction with tunnel resistance $G^{-1}_{\text{T}}= 270 \, \mathrm{k} \Omega$
is embedded in a $\lambda /4$ coplanar resonator whose inner
conductor is made of an array of identical and equally spaced Al/AlO$_\mathrm{x}$/Al
SQUIDs.  The SQUIDs' Josephson inductance can be increased by an external magnetic flux, increasing the resonator's characteristic impedance $Z_{C}$ from 1.2 k$\Omega$ to 1.8 k$\Omega$, while decreasing its resonant frequency $\nu_0$ from 6 GHz to 4 GHz \cite{AltimirasHighZMatchAPL2013, AltimirasNoiseDCBPRL2014}. A 30~$\times$~50~$\times$~0.3~$\mu$m\tmrsup{$3$} gold patch is inserted between the tunnel junction and the SQUID array to evacuate the Joule power generated
at the tunnel junction via electron-phonon coupling. The chip is connected to
a commercial $50\,\Omega$ matched bias tee. The low frequency path is used to
bias the sample through a cold $13 \,\mathrm{M} \Omega$
resistor, and to measure the tunnel junction dc conductance $G(V_{dc}, \alpha)=dI/dV_{dc}$. The RF path is connected to a 20~dB directional coupler, which allows us to (i) shine microwave power on the tunnel junction through the weakly coupled port while (ii) measuring the microwave signals emitted/reflected by the sample via the well transmitted port, itself
connected to a cryogenic amplifier with a $\sim$2.5~K noise
temperature in the 4-8~GHz bandwidth. Two cascaded circulators (only one being represented) divert the back-action noise of the amplifier onto thermalized $50\,\Omega$ matched loads.  The low temperature $k_\mathrm{B} T \ll h\nu_0$ and the high tunnel resistance  $G_T ^{-1} = 270\,\mathrm{k}\Omega$ ensure negligible photon occupation in the resonator \cite{SM}.

The dc-conductance of the junction is deduced from the voltage drop across the sample induced by a sinusoidal $5\,\mathrm{nA}_{\mathrm{RMS}}$ current modulation at 12~Hz through the low frequency port, measured via homodyne detection. The $I\left(V_{dc}\right)$ curve is then calculated by numerical integration of the differential conductance shown in Fig. \ref{Figure2}, which is in good agreement with the DCB-theory prediction based on our microwave design \cite{AltimirasNoiseDCBPRL2014}. This allows us to use the designed impedance seen by the junction, $Z\left(\nu\right)$ to extract the admittance of the tunnel junction from the microwave signal reflected by the sample. 

More specifically, we inject a small coherent tone delivered by a Vectorial Network Analyzer (VNA) at the resonant frequency tuned to $\nu_0=4.1 GHz$, into the resonator through the -20~dB port of the directional coupler. The reflected signal is then amplified and sent to the VNA input port. The $\sim$ -140 dBm excitation signal amplitude yields a $\sim 1.4 \mu V_{\mathrm{RMS}}$ ac voltage on the sample, corresponding to $\alpha \simeq 0.1$, making photo-assisted tunneling negligible, thus ensuring a linear response. Due to the finite $\sim 15\,\mathrm{dB}$ directivity of the coupler, a coherent leak adds up to the signal, so that the total transmission coefficient can be expressed as $S_{\mathrm{out,in}}(\nu_0)=G\left(\Gamma+F\right)$ where $\Gamma$ stands for the reflection coefficient at the input of the resonator, $F$ the coherent leak transmission, and $G$ stands for the total gain of the chain (including the attenuation of the various microwave components). The gain $G$ and the leakage coefficient $F$ can be calibrated by two measurements of transmission $S_{\mathrm{out,in}}$: (i) we first apply a large dc bias $eV \gg h\nu_0$  to the sample and assume that the corresponding junction admittance is given by the tunnel conductance, $\lim\limits_{V_{dc}\to \infty} Y(\nu,V_{dc})=G_T$ \cite{SM}, and then (ii) detune the resonator frequency to $\sim 3.7\,\mathrm{GHz}$, ensuring that $\Gamma \simeq-1$ \cite{SM}.  We then measure $S_{21}(\nu_0)$ as a function of the bias voltage, from which we extract the variations of the finite frequency admittance. As shown in Fig. \ref{Figure2}, the data are in good agreement with our theoretical predictions: the junction's conductance $\operatorname{Re}[Y(\nu_0,V_{dc},\alpha=0)]$ is well described by Eq. \ref{Ydenu} and its susceptance $\operatorname{Im}[Y(\nu_0,V_{dc},\alpha=0)]$ is negligible.    

In a second experiment, we measure the derivative of the shot noise spectral density  $\partial S_I (\nu_0,V_{dc},\alpha=0)/\partial V_{dc}$ at the resonant frequency with respect to the dc bias in absence of an RF-drive \cite{AltimirasNoiseDCBPRL2014}: the output of the amplifying chain is connected to a 180~MHz room temperature adjustable filter centered around $\nu_0$ and to a quadratic detector whose output voltage is proportional to the noise power. We perform an homodyne detection of the variations of the system noise temperature induced by the 12~Hz modulation used to measure the conductance. Due to the impedance mismatch between the admittance of the sample and the impedance of the rf-detection chain $|Y(\nu,V_{dc})Z(\nu)|\ll 1$, the emitted power density reads $S_I\left(\nu,V_{dc},\right)\operatorname{Re}Z\left(\nu\right)/\left|1+Y(\nu,V_{dc})Z\left(\nu\right)\right|^2$. The voltage dependence of the coupling coefficient arising from Coulomb blockade,  $\sim 1\,\%$, can be neglected, so that we extract directly $\partial S_I (\nu_0,V_{dc})/\partial V_{dc}$ from the noise temperature modulations.  The FDT relation Eq. (\ref{FDT_I-EmissionNoise}) is in agreement with the experimental results shown in Fig. \ref{Figure2} . 

For the photo-assisted experiments, a rf drive, $\sim -120\,\mathrm{dBm}$ at $\nu_0-5$ MHz, is superposed to the VNA signal using a room temperature -10dB  directional coupler, as shown in Fig. \ref{Figure1}. The induced coherent state amplitude $\alpha$ at the input of the tunnel junction can be estimated from the independently calibrated attenuation of the feed line, and from the voltage divider $\left[1+Z(\nu_0)\overline Y(\nu_0,V_{dc},\alpha)\right]^{-1}$. The experimental $\overline G(V_{dc},\alpha)$ data shown in Fig. \ref{Figure3} a) are well reproduced by the Tien-Gordon relation Eq. (\ref{eq:I-Tien-Gordon}) using $\alpha$ as a fitting parameter. The extracted value $\alpha=1.15$, is in agreement with the estimated value within 15\% (1 dB). Moreover since $Z(\nu_0)Y(\nu_0,V_{dc}) \ll 1$, the variations of $Y(\nu_0,V_{dc})$ with $V_{dc}$ induce negligible variations of $\alpha$ with the dc bias. The driving frequency is chosen close, but different than the VNA frequency, so that $Y(\nu\simeq\nu_0, V_{dc}, \alpha)$ can still be measured. As shown in Fig. \ref{Figure3} b) it follows a photo-assisted relation and hence Eq. (\ref{Ydenu}) is well obeyed.

For the emission shot noise power measurement $\overline{S_I}(\nu,V_{dc},\alpha)$, we eliminate the driving tone parasitic signal by implementing a band rejection filter: We mix the total signal with a reference at the driving frequency, and low pass filter the down converted signal with a 80 MHz low pass filter, which is then fed to the quadratic detector. Thanks to the 1 MHz low frequency cut-off of the quadratic detector,
 its output is insensitive to the driving tone reflected signal. The results, shown in the right panel of Fig. \ref{Figure3} are found to follow the photo-assisted relation Eq. \ref{eq:SI-Tien-Gordon}, so that our FDT relation Eq. \ref{FDT_I-EmissionNoise} also holds for a time dependent bias.

In conclusion, we have shown theoretically and experimentally that in the presence of strong quantum fluctuations of the driving voltage, the finite frequency admittance and current fluctuations of a tunnel element follow fluctuation-dissipation relations derived for classical drives. This also holds in the presence of a time dependent bias, where photo-assisted tunneling expressions are also valid. Our derivation relies on the fact that no memory effect occurs neither in the electromagnetic environment nor in the quantum conductor, and that both follow a detailed balance relation. Our experimental approach is very general and can be readily exploited to test fluctuation-dissipation relations for systems not fulfilling our hypothesis, for instance for conductors beyond the weak coupling limit like Quantum Point Contacts \cite{PhysRevLett.86.4887,PhysRevLett.87.046802,AltimirasNoiseDCBPRL2014,ParmentierStrongDCB2011,PhysRevB.88.205419}, where DCB was recently demonstrated to bear a connection to the physics of impurities in Luttinger liquids {\cite{JezouinStrongDCB2013}},  or in systems having rich internal dynamics such as Quantum Dots \cite{SukhoLossCotunnNoisePRB2001,MebrahtuNature488p61,PhysRevLett.108.046802}. 

\section{Acknowledgements}
We gratefully acknowledge support from the CNano-IDF Shot-E-Phot and Masquel, the Triangle de la Physique DyCoBloS and
ANR AnPhoTeQ, and the CNR COCA grants.

\balancecolsandclearpage

\begin{center}
\textbf{\large{Supplemental Materials: Fluctuation-dissipation relations of a tunnel junction driven by a quantum circuit}} 
\end{center}
\setcounter{equation}{0}
\setcounter{figure}{0}
\setcounter{table}{0}
\setcounter{page}{1}
\makeatletter
\renewcommand{\theequation}{S\arabic{equation}}
\renewcommand{\thefigure}{S\arabic{figure}}

This supplementary material provides the complete theoretical derivation of the formulas provided in the article body, insisting on the origin of the detailed balance relations allowing to derive the fluctuation-dissipation relations linking the current fluctuations to the current-voltage $I(V)$ characteristic of the junction, modified by its environment. It also contains the full details regarding the experimental procedure used to extract the real and imaginary part of the tunnel junction's admittance.
	
\section{Theoretical derivation}
\subsection{Defining the problem}

\begin{figure}[t]
  \includegraphics[width=8cm]{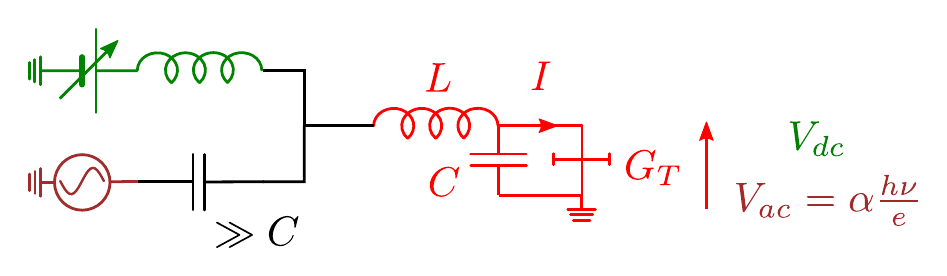}
  \caption{Model system: a small conductance tunnel junction $G_T$ is embedded in a resonator of frequency $\nu_0=1/\sqrt{L C}$ and characteristic impedance $Z_C=\sqrt{L/C}$. The system is voltage connected to a dc (rf) voltage source $V_{dc}$ ($V_{ac}$) through a large inductance (capacitance).}
	\label{FigureS1}
\end{figure}

The circuit we deal with is that of a tunnel junction shunted by a harmonic oscillator. As depicted in Fig. \ref{FigureS1} the tunnel element sees a resonant circuit of resonant frequency $\nu_0=\frac{1}{2\pi\sqrt{LC}}$, and characteristic impedance $Z_c=\sqrt{\frac{L}{C}}$, where $C$ is the oscillator's capacitance (including tunnel junction self capacitance), and $L$ its inductance. The system is dc biased through a large inductance and an ac drive at the resonant frequency is applied through a large capacitance. We describe this circuit by a Hamiltonian consisting in the sum of three terms \cite{ingoldnazarov1992DCB} $H = H_{qp} + H_{env} + H_T$. The first $H_{qp}=\sum_{l}\epsilon_l n_l +\sum_r \epsilon_r n_r$, with $n_{l,r}=c_{l,r}^\dagger c_{l,r}$ being the occupation number of the fermionic quasi-particle operators, describes the (free) quasi-particle dynamics (in the sense of the Landau theory of an interacting Fermi sea \cite{PinesNozieres}) at the left and right electrodes of the tunnel junction. The second term $H_{env}=\frac{Q^2}{2 C}+\frac{\Phi^2}{2L}$ describes the dynamics of the LC resonator in terms of the conjugated electromagnetic variables $Q$, the influence charge at the plates of the tunnel junction, and $\Phi$ the magnetic flux stored in the inductance which is related to the voltage  drop $V$ across the junction as $\Phi(t)=\int_{-\infty}^{t}\mathrm{d}t'V(t')$. The last term $H_T=T+T^\dagger$, where $T=e^{ie\Phi/\hbar} \Theta $ with $\Theta = \sum_{l,r}\tau_{lr}c_l^\dagger c_r$, describes the tunnel coupling which transfers quasi-particles between both electrodes with the (small) probability amplitudes $\tau_{l, r}$ and accounts for the corresponding charging of the capacitance since $e^{-ie\Phi/\hbar}Qe^{ie\Phi/\hbar}=Q-e$. $H_T$ is the minimal electrodynamic coupling of the quantum conductor to its electromagnetic environment. It is valid in the long wavelength limit with respect to the size of the electrodes \cite{LebedevPhotonStatsPRB2010}, neglecting their intrinsic electrodynamics \cite{PierreLDOSPRL2001} beyond the mean-field approximation encompassed in the shunting capacitance. The coupling Hamiltonian $H_T$ correctly treats the charge accumulated at the electrodes by conserving the current at the node represented by the red dot in Fig.~1: Defining the quasi-particle current as $I_{qp}=e\dot{n_l}=-\frac{ie}{\hbar}[n_l, H]=-\frac{ie}{\hbar}(T-T^\dagger)$, the displacement current through the capacitance as $I_D=\dot{Q}=-\frac{i}{\hbar}[Q, H]=\frac{ie}{\hbar}(T-T^\dagger)+\Phi/L$, and the current flowing through the inductance $I_L=\Phi/L$, one finds $I_{qp}+I_D=I_L$ which is an exact equality for the current operators.

All the existing literature deriving fluctuation-dissipation relations in tunnel junctions \cite{DahmJosephsonLinewidthPRL1964,RogovinScalapino1974,LeeLevitovDCBPRB1996,SukhoLossCotunnNoisePRB2001,PhysRevLett.96.136804,Safi-Joyez,SafiTunnel} where derived considering the bias is an eventually time-dependent classical external parameter. Here we exploit the dynamical Coulomb blockade formalism introduced above to provide a full quantum mechanical description of the time dependent bias. To describe the field in the resonator under a coherent drive at the resonant frequency $\nu_0$, we assume that the density matrix of the resonator corresponds to a "displaced thermal state" \cite{doi:10.1080/09500348814550571}: $\rho_{env}=D(i\alpha/2r)\rho_T D(i\alpha/2r)^\dagger$. Here, $\rho_T=\exp(-\beta H_{env})/\mathrm{Tr}(\exp(-\beta H_{env}))$ is the usual density matrix for the resonator at temperature $\beta^{-1}$, $r=\sqrt{\frac{\pi Z_c}{R_K}}$. This choice is justified by the resulting quantum average time dependent voltage across the junction: $\mathrm{Tr}(\rho_{env}\dot{\Phi})=V_{ac}\cos(2\pi \nu_0 t)$ for $\alpha=eV_{ac}/h\nu_0$. We further assume that the resonator's state is unperturbed by the tunneling events. Our approach takes into account the thermal fluctuations of the environment, and allows us to compute non-only the dc current as done in \cite{SouquetNonCPATarxiv}, but we compute also its time dependence, the admittance and the power spectral density of current fluctuations using standard perturbation techniques to lowest order.

Making use of the interaction picture of the current operator $I_0(t)$ with respect to the uncoupled evolution $H_0=H_{qp}+H_{env}$, its full time evolution up to first order in the tunnel coupling reads:
\begin{widetext}
\begin{align*}
I(t)=&I_0(t)-\frac{i}{\hbar}\int_{-\infty}^{t}[H_C(t'),I_0(t)]\mathrm{d}t'\\
=&\frac{-ie}{\hbar}(T(t)-T^\dagger(t))-\frac{e}{\hbar^2}\int_{-\infty}^{t}[T(t')+T^\dagger(t'), T(t)-T^\dagger(t)]\mathrm{d}t'.
\end{align*}
\end{widetext}

In the following, we will have to calculate quantum average of various operators, which are meant to be taken with respect to the original states, described by the factorized density matrix of uncoupled thermal quasi-particles and the displaced thermal environment \cite{doi:10.1080/09500348814550571}: 
\begin{equation*}
\rho_0=\frac{\exp(-\beta H_{qp})}{\mathrm{Tr}(\exp(-\beta H_{qp}))}\otimes \frac{D(i\alpha/2r)\exp(-\beta H_{env})D(i\alpha/2r)^\dagger}{\mathrm{Tr}(\exp(-\beta H_{env}))}.
\end{equation*}

Since the operator $T$ does not conserve the quasi-particle number, the non-interacting mean current vanishes $\langle I_0(t)\rangle=0$, and the evaluation of the mean current must be kept to first order in the tunneling coupling, where the non-vanishing terms read:
\begin{align}\label{eq:MeanCurrent}
\langle I(t)\rangle=\frac{2e}{\hbar^2}\operatorname{Re}\int_0^{+\infty}\mathrm{d}\tau\langle T(t+\tau)T^\dagger(t)\rangle-\langle T^\dagger(t+\tau)T(t)\rangle.
\end{align}
The average of the current correlations is already finite at zeroth order, and reduces to the two only quasi-particle number conserving terms:
\begin{align}\label{eq:TunnelTimeCorre}
\langle I(t+\tau)I(t)\rangle=\frac{e^2}{\hbar^2}\big(\langle T(t+\tau)T^\dagger(t)\rangle +\langle T^\dagger(t+\tau)T(t)\rangle\big).
\end{align}
Therefore, the problem reduces to compute two correlations functions $\langle T(t+\tau) T(t)^\dagger\rangle$ and $\langle T^\dagger (t+\tau)T (t)\rangle$.

\subsection{Correlation functions}
Since the tunnel coupling $T(t)=e^{ie\Phi(t)/\hbar}\Theta(t)$ is factorized in terms of environment and quasi-particle operators, and so is is the density matrix, the correlation functions are also factorized: $\langle T(t+\tau) T(t)^\dagger\rangle=\langle\Theta(t+\tau)\Theta^\dagger(t)\rangle\langle e^{ie\Phi(t+\tau)/\hbar}e^{-ie\Phi(t)/\hbar}\rangle$ and $\langle T^\dagger(t+\tau) T(t)\rangle=\langle\Theta^\dagger(t+\tau)\Theta(t)\rangle\langle e^{-ie\Phi(t+\tau)/\hbar}e^{ie\Phi(t)/\hbar}\rangle$.

\subsubsection{Quasi-particle correlation functions}
Introducing the density of quasi-particle states on the left and right electrodes $\rho_{l, r}(\epsilon)$, and their mean occupation number $f_{l,r}(\epsilon)=(1+\exp(-\beta_{l,r}\epsilon))^{-1}$ we have:
\begin{widetext}
\begin{align*}
\langle \Theta(t+\tau)\Theta^\dagger(t) \rangle =& \sum_{l,r} |\tau_{lr}|^2 \langle c_l^\dagger(t+\tau) c_l(t)\rangle \langle c_r(t+\tau)c_r^\dagger(t) \rangle \\
=&  \int \mathrm{d}\epsilon\mathrm{d}\epsilon ' |\tau(\epsilon,\epsilon')|^2 \rho_l(\epsilon) \rho_r(\epsilon') f_l(\epsilon) (1 - f_r(\epsilon') ) e^{-i(\epsilon-\epsilon')\tau/\hbar} \\
=&\int \mathrm{d}\epsilon  \mathrm{d}\epsilon '|\tau(\epsilon,\epsilon+\epsilon')|^2  \rho_l(\epsilon) \rho_r(\epsilon+\epsilon') f_l(\epsilon) (1 - f_r(\epsilon+\epsilon') ) e^{-i\epsilon'\tau/\hbar} \\
=& \int  \mathrm{d}\epsilon ' \theta(\epsilon') e^{-i\epsilon'\tau/\hbar} \\
=& \theta (\tau)
\end{align*}
\end{widetext}
where we defined the function $\theta (\epsilon')=\int\mathrm{d}\epsilon|\tau(\epsilon,\epsilon+\epsilon')|^2 \rho_l(\epsilon) \rho_r(\epsilon+\epsilon') f_l(\epsilon) (1 - f_r(\epsilon+\epsilon'))$ counting the number of all possible inelastic quasi-particle transfers from the left electrode to the right electrode with an energy difference $\epsilon'$. The correlation function $\langle \Theta(t+\tau)\Theta^\dagger(t) \rangle $ only depends on the time difference $\tau$, a consequence of averaging it over stationary (equilibrium) states.

Similarly , we obtain for the second quasi-particle term:

\begin{widetext}
\begin{align*}
\langle \Theta^\dagger( t +\tau)\Theta(t) \rangle =& \sum_l |\tau_{lr}|^2 \langle c_r^\dagger(t+\tau) c_r(t)\rangle \langle c_l(t+\tau)c_l^\dagger(t) \rangle \\
=&  \int \mathrm{d}\epsilon \mathrm{d}\epsilon ' |\tau(\epsilon',\epsilon)|^2\rho_r(\epsilon) \rho_l(\epsilon') f_r(\epsilon) (1 - f_l(\epsilon') ) e^{-i(\epsilon-\epsilon')\tau/\hbar} \\
=&  \int \mathrm{d}\epsilon \mathrm{d}\epsilon '|\tau(\epsilon+\epsilon',\epsilon)|^2 \rho_r(\epsilon) \rho_l(\epsilon+\epsilon') f_r(\epsilon) (1 - f_l(\epsilon+\epsilon') ) e^{-i\epsilon'\tau/\hbar} \\
=&  \int  \mathrm{d}\epsilon '\theta^*(\epsilon') e^{-i\epsilon'\tau/\hbar} \\
=& \theta^\ast (\tau).
\end{align*}
\end{widetext}
where now, the function $\theta^\ast (\epsilon')=\int\mathrm{d}\epsilon|\tau(\epsilon+\epsilon',\epsilon)|^2 \rho_r(\epsilon) \rho_l(\epsilon+\epsilon') f_r(\epsilon) (1 - f_l(\epsilon+\epsilon'))$ counts the number of all possible inelastic quasi-particle transfers from the right electrode to the left electrode with an energy difference $\epsilon'$.
If both electrodes are fully symmetric, namely $\rho_l=\rho_r$ and $f_l=f_r$, and $|\tau_{l,r}|^2=|\tau_{r,l}|^2$ it is evident that $\theta=\theta^*$. However, exploiting the identity $f_{l, r}(-\epsilon)=1-f_{l, r}(\epsilon)$, one can see that $\theta=\theta^*$ still holds provided the system has has an electron-hole symmetry $\rho_{l,r}(\epsilon)=\rho_{l,r}(-\epsilon)$ and $|\tau(\epsilon,\epsilon')|^2=|\tau(-\epsilon,-\epsilon')|^2$. Physically, it means the system does not give rise to any thermoelectric effect. In the following to simplify our calculations, we will assume that we have such an electron-hole symmetric system which is valid for normal NIN junction, superconducting $\mathrm{S}_1\mathrm{IS}_2$ junctions (allowing different gaps) or hybrid NIS junctions, in the experimentally relevant limit of small energies with respect to the Fermi energy and barrier height. Therefore we have $\langle \Theta(t+\tau)\Theta(t)^\dagger \rangle =\langle \Theta^\dagger(t+\tau)\Theta(t) \rangle=\theta(\tau)$, which, as shown in the section on detailed balance relations below, it automatically ensures that $\theta(\epsilon)$ follows a detailed balance relation, namely $\theta(\epsilon)=e^{-\beta \epsilon}\theta(-\epsilon)$. However, the general detailed balance symmetry Eq. \ref{A1} allows to derive the same results without assuming a particle-hole symmetry. The key point being that the Fourier transform of $\theta(t)$ and $\theta^\ast(t)$ are related by a detailed balance symmetry, which is valid as soon as they are the result of averaging over a thermal equilibrium state (see section on detailed balance relations below) \cite{CallenWelton,SukhoLossCotunnNoisePRB2001,SafiTunnel}.

\subsubsection{Environment correlation functions}
We fist recall how the normalized quantum fluctuations of the flux operator are recast in terms of the normalized modes of the LC resonator (the dc voltage does not modify the dynamics of the resonator, see e.g. \cite{ingoldnazarov1992DCB}):
\begin{align*}
\delta \phi(t)=e\Phi(t)/\hbar-eV_{dc}t/\hbar=r(a^\dagger(t)+a(t)),
\end{align*}
with $r=\sqrt{\frac{\pi Z_c}{R_K}}$. This relation enables to express the operators $\exp[\pm i\delta\phi(t)]$, and thus the correlation functions, in terms of a displacement operator:
\begin{align*}
\exp[\pm i\delta\phi(t)]=&\exp[\pm (ire^{2i\pi \nu_0 t}a^\dagger +ire^{-2i\pi \nu_0 t}a)]\\
=&D[\pm ire^{2i\pi \nu_0 t}].
\end{align*}
Then we note, thanks to the Campbell-Baker-Hausdorff identity, that displacement operators have the following commutation:
\begin{align*}
D[\alpha]D[\beta]=e^{\alpha\beta^\ast-\alpha\beta^\ast}D[\beta]D[\alpha].
\end{align*} And finally, we exploit this algebra, the invariance of trace with respect to cyclic permutations, and that displacement operators are unitary $D[\alpha]D[\alpha]^\dagger=1$, to simplify the correlation functions as:

\begin{widetext}
\begin{align*}
\langle \exp[\pm ie\Phi(t+\tau)&/\hbar]\exp[\mp ie\Phi(t)/\hbar] \rangle\\
=&e^{\pm ieV_{dc}\tau/\hbar}\mathrm{Tr}(D[i\alpha/2r]\rho_\beta D[i\alpha/2r]^\dagger \exp[\pm i\delta\phi(t+\tau)]\exp[\mp i\delta\phi(t)])\\
=&e^{\pm ieV_{dc}\tau/\hbar}\mathrm{Tr}(\rho_\beta D[i\alpha/2r]^\dagger D[\pm ire^{2i\pi \nu_0 (t+\tau)}]D[\mp ire^{2i\pi \nu_0 t}]D[i\alpha/2r])\\
=&e^{\pm ieV_{dc}\tau/\hbar}\mathrm{Tr}(\rho_\beta D[\pm ire^{2i\pi \nu_0 (t+\tau)}]D[\mp ire^{2i\pi \nu_0 t}])e^{\pm i\alpha \sin(2\pi \nu_0 (t+\tau))}e^{\mp i\alpha \sin (2\pi \nu_0 t)}\\
=&e^{\pm ieV_{dc}\tau/\hbar}\mathrm{Tr}(\rho_\beta \exp[\pm i\delta\phi(t+\tau)]\exp[\mp i\delta\phi(t)])e^{\pm i\alpha \sin(2\pi \nu_0 (t+\tau))}e^{\mp i\alpha \sin (2\pi \nu_0 t)}\\
=& e^{\pm ieV_{dc}\tau/\hbar}e^{J(\tau)}e^{\pm i\alpha \sin(2\pi \nu_0 (t+\tau))}e^{\mp i\alpha \sin (2\pi \nu_0 t)}.
\end{align*}
\end{widetext}
In the last equation we identified the standard (stationary) correlation function found in dynamical Coulomb blockade theory \cite{ingoldnazarov1992DCB}, $e^{J(\tau)}=\mathrm{Tr}(\rho_\beta \exp[\pm i\delta\phi(t+\tau)]\exp[\mp i\delta\phi(t)])$, which is the Fourier transform of the so-called $P(E)$ function weighting the probability for a tunneling event to exchange the amount of energy $E$ with the resonator. It can be shown (see  \cite{ingoldnazarov1992DCB} and the section on detailed balance relations below), that $P(E)$ obeys a detailed balance relation. It is noteworthy that the time-dependent phases resulting from the action of the displacement operators into the coupling operator are exactly those one would obtain for a semi-classical treatment where the time-dependent bias is treated as a classical parameter. At a technical level, this is why we obtain the same results as those derived for a classical drive.

\subsubsection{Summing up}

Finally we pick all the terms and, exploiting the Jacobi-Angers expansion of the time-dependent exponentials in terms of Bessel functions of the first kind, we obtain:
\begin{widetext}
\begin{align}\label{eq:corrFuncs}
\langle T(t+\tau) T(t)^\dagger\rangle=& \sum_{k,l}J_k(\alpha)J_l(\alpha)e^{i(eV_{dc}+ kh\nu_0)\tau/\hbar}e^{-2i\pi (k-l)\nu_0t}\theta(\tau)e^{J(\tau)}\\
\langle T^\dagger (t+\tau)T (t)\rangle=&\sum_{k,l}J_k(\alpha)J_l(\alpha)e^{-i(eV_{dc}+ kh\nu_0\tau)/\hbar}e^{+2i\pi (k-l)\nu_0t}\theta(\tau)e^{J(\tau)}.
\end{align}
\end{widetext}

\subsection{Time-dependent mean current}

Inserting back the correlations functions Supp. Eqs. (4-5) into the expression for the mean current Supp. Eq. (1) we have:
\begin{widetext}
\begin{align*}
\langle I(t)\rangle=\frac{2e}{\hbar^2}\operatorname{Re}\int_0^{+\infty}\mathrm{d}\tau \theta(\tau)e^{J(\tau)}\sum_{k,l}J_k(\alpha)J_l(\alpha)(e^{i(eV_{dc}+kh\nu_0)\tau/\hbar}e^{-2i\pi(k-l)\nu_0t}-c.c.)
\end{align*}
\end{widetext}
Fourier transforming this time dependence, we obtain a non-zero response only for the harmonics of the driving field frequency:
\begin{widetext}
\begin{align*}
I(\Omega)=\delta(\Omega-n\nu_0)\frac{2e}{\hbar^2}\operatorname{Re}\int_0^{+\infty}\mathrm{d}\tau \theta(\tau)e^{J(\tau)}\sum_{k}J_k(\alpha)J_{k+n}(\alpha)(e^{i(eV_{dc}+kh\nu_0)\tau/\hbar}-c.c.).
\end{align*}
\end{widetext}
All the dynamical response of the mean current can thus be reconstructed from the characteristic obtained under a stationary bias $I(V_{dc})$:
\begin{widetext}
\begin{align*}
I(\Omega)=&\delta(\Omega-n\nu_0)\sum_{k}J_k(\alpha)J_{k+n}(\alpha)\frac{2e}{\hbar^2}\operatorname{Re}\int_0^{+\infty}\mathrm{d}\tau \theta(\tau)e^{J(\tau)}(e^{i(eV_{dc}+kh\nu_0)\tau/\hbar}-c.c.)\\
=&\delta(\Omega-n\nu_0)\sum_{k}J_k(\alpha)J_{k+n}(\alpha) I(eV_{dc}+kh\nu_0).
\end{align*}
\end{widetext}

\subsubsection{Photo-assisted relation for the current}

In particular we find that the time-averaged current is provided by a photo-assisted (or Tien-Gordon) relation \cite{TienGordonPR1963,SafiSukho}:

\begin{align*}
\overline{\langle I(t)\rangle }=I(\Omega=0)=\sum_k J_k(\alpha)^2 I(eV_{dc}+kh\nu_0),
\end{align*}

which is Eq. 1 of the main text.

\subsection{Admittance}

\subsubsection{Non-linear current response}

We can also compute the in-phase response for all the harmonics, which can be formally expressed as $I_{X1}^n=I(n\nu_0)+I(-n\nu_0)$:
\begin{widetext}
\begin{align*}
I_{X1}^n=&\sum_{k}\Big(J_k(\alpha)J_{k+n}(\alpha)+J_k(\alpha)J_{k-n}(\alpha)\Big) I(eV_{dc}+kh\nu_0)\\
=&\sum_k J_k(\alpha)J_{k+n}(\alpha) \Big(I(eV_{dc}+kh\nu_0)+(-1)^n I(eV_{dc}-kh\nu_0)\Big)\\
\end{align*}
\end{widetext}
where we exploited the symmetry of Bessel functions of the first kind $J_{-k}(\alpha)=(-1)^kJ_{k}(\alpha)$.

The out-of-phase response $I_{X2}^n$ is more tedious, but one arrives to the result:
\begin{align*}
I_{X2}^n=-\frac{1}{\pi} \mathcal{P}\int \mathrm{d}\nu'\frac{I_{X1}^n(V_{dc}, f')}{f'-\nu_0}
\end{align*}
which is nothing but a Kramers-Kronig relation between the in- and out-of phase responses to all the harmonics. This was expected since the current is a physical observable: causal and real valued (as is directly visible in Supplementary Eq.~(\ref{eq:MeanCurrent})).

\subsubsection{Low bias limit}

The stationary admittance probed at the resonator frequency is now straightforward: exploiting the asymptotic form of Bessel functions $J_k(\alpha)\simeq \frac{1}{k!}(\frac{\alpha}{2})^k$ valid for $\alpha\ll 1$ and retaining only first order terms in $V_{ac}$ we find:

\begin{align*}
\operatorname{Re}Y(\nu_0)=\lim_{\alpha\ll 1} \frac{I_{X1}^1}{V_{ac}}=e\frac{I(eV_{dc}+h\nu_0)-I(eV_{dc}-h\nu_0)}{2h\nu_0},
\end{align*}
which is Eq. 3 of the main text.  The imaginary part of the junction's admittance $\operatorname{Im}Y(\nu_0)$ follows from Kramers-Kronig relations.

\subsubsection{Photo-assisted relation for the admittance}

In order to derive the admittance at other frequencies, and to allow having an independent pumping as is done in the experiment, we introduce a second oscillator of resonant frequency $f_1$ coupled to the junction. The trick is that defining it with a vanishing characteristic impedance, it does not give rise to any back-action to the tunnel junction (namely $e^{J(\tau)}$ is unchanged), yet it allows to drive it at arbitrary frequencies with an amplitude $\alpha_1$. As a result we have a time-dependence resulting from the beating of these two sources, which we have taken with the same phase to ease notations:

\begin{widetext}
\begin{align*}
\langle I(t)\rangle=\frac{2e}{\hbar^2}\operatorname{Re}\int_0^{+\infty}\mathrm{d}\tau &\theta(\tau)e^{J(\tau)}\sum_{k,l,m,n}J_k(\alpha)J_l(\alpha)J_m(\alpha_1)J_n(\alpha_1)\times\\
&(e^{i(eV_{dc}+kh\nu_0+mhf_1)\tau/\hbar}e^{-2i\pi(k-l)\nu_0t}e^{-2i\pi(m-n)f_1 t}-c.c.)
\end{align*}
\end{widetext}

Now we take the in-phase response with respect to the first harmonic of the vanishing mode:

\begin{widetext}
\begin{align*}
\frac{2e}{\hbar^2}\operatorname{Re}\int_0^{+\infty}\mathrm{d}\tau \theta(\tau)e^{J(\tau)}\sum_{k,l,m}J_k(\alpha)J_l(\alpha)J_m(\alpha_1)&(J_{m+1}(\alpha_1)+(J_{m-1}(\alpha_1))\times\\
&(e^{i(eV_{dc}+kh\nu_0+mhf_1)\tau/\hbar}e^{-2i\pi(k-l)\nu_0 t}-c.c)
\end{align*}
\end{widetext}

and we average over time this quadrature to obtain another photo-assisted relation for the in-phase quadrature:
\begin{widetext}
\begin{align*}
\sum_{k}J_k(\alpha)^2\sum_m J_m(\alpha_1)J_{m+1}(\alpha_1) \Big(I(eV_{dc}+kh\nu_0+mhf_1)-I(eV_{dc}+kh\nu_0-mhf_1)\Big).
\end{align*}
\end{widetext}

Finally, taking the limit $\alpha_1\ll 1$ we obtain the photo-assisted relation for the real part of the admittance at arbitrary frequency $f_1$, in the presence of an arbitrary coherent pumping at frequency $\nu_0$ with amplitude $\alpha=eV_{ac}/h\nu_0$  \cite{SafiTunnel}:
\begin{align*}
\overline{\operatorname{Re}Y(V_{dc},f_1, \alpha)}=\sum_k J_k(\alpha)^2 \operatorname{Re}Y(eV_{dc}+kh\nu_0,f_1, \alpha=0).
\end{align*}

\subsection{Current fluctuations}
Inserting back the expression of the correlation functions Supplementary Eqs. (3-4) into the current fluctuation Supplementary Eq.~(\ref{eq:TunnelTimeCorre}) we find:
\begin{widetext}
\begin{align*}
\langle I(t+\tau)I(t)\rangle=&\frac{e^2}{\hbar^2}\big( \theta(\tau)e^{J(\tau)}\sum_{k,l}J_k(\alpha)J_l(\alpha)(e^{i(eV_{dc}+kh\nu_0)\tau/\hbar}e^{-2i\pi(k-l) \nu_0 t}+c.c.)  \big).
\end{align*}
\end{widetext}

Again, the Fourier transform of the time ($t$) dependence of the current time ($\tau$) correlations can be fully expressed as copies of the stationary correlations $\langle I(\tau)I(0)\rangle_0$ arising for a stationary bias:
\begin{widetext}
\begin{align*}
\mathcal{FT}\langle I(t+\tau)I(t)\rangle_{[\Omega]}=\delta(\Omega-nh\nu_0)\sum_{k}J_k(\alpha)J_{k+n}(\alpha) \mathcal{FT} \langle I(\tau)I(0)\rangle_0[eV_{dc}+kh\nu_0].
\end{align*}
\end{widetext}

One recognizes the same structure as the one we obtained for the mean current harmonics. Therefore, the mean current, and the mean current fluctuations have exactly the same time dependence. This means that a quantum regression theorem applies to the system \citep{FordOConnellRegrThmPRL1996}. Which is a consequence of the stochastic nature of tunneling events: no memory effects build neither in the quasi-particle nor in the environment. Eq. 2 of the main text, corresponding to the time averaged emission noise current density, is obtained for $n=0$.

Finally, since the current time correlation has the same formal dependence as the mean current, one automatically recovers the same photo-assisted relations not only for the current time correlations, but also for their power spectral density which is measured in the experiment. We also recover the "noise susceptibility" found, and measured, in \cite{GabelliNoiseSensiPRL2008}, for the in-phase response of the power density of current fluctuations. We stress there is nothing genuine to the noise, since the mean current has the same structure.

\section{Detailed balance relations}\label{AppendixDetailBalance}

Any equilibrium time correlation $\langle A(t)B(0)\rangle$ and $\langle B(0)A(t)\rangle$ follow a detailed balance relation, that is their Fourier transforms are related as:
\begin{align}
\mathcal{FT}[\langle A(t)B(0)\rangle]_{(\nu)}=e^{-\beta h\nu}\mathcal{FT}[\langle B(0)A(t)\rangle]_{(\nu)}.
\label{A1}
\end{align}
This results from the invariance of trace under cyclic permutation:
\begin{align*}
\langle A(t)B(0)\rangle=&Tr\Big(e^{-\beta H}e^{iHt/\hbar}Ae^{-iHt/\hbar}B\Big)/Z\\
=& Tr\Big(Be^{-\beta H}e^{iHt/\hbar}Ae^{-iHt/\hbar}e^{+\beta H}e^{-\beta H}\Big)/Z\\
=& Tr\Big(e^{-\beta H}Be^{iH(t+i\hbar\beta)/\hbar}Ae^{-iH(t+i\hbar\beta)/\hbar}\Big)/Z\\
=&\langle B(0)A(t+i\hbar\beta)\rangle,
\end{align*} 
 and the translation theorem for Fourier transforms.
 
 In the particular case of autocorrelations $(A=B)$, one immediately  obtains:
$S_A(\nu)=e^{-\beta h\nu}S_A(-\nu)$,
since 
\begin{align*}
S_A(-\nu)=&4\pi\mathcal{FT}[\langle A(t)A(0)\rangle]_{(-\nu)}\\
=&4\pi\mathcal{FT}[\langle A(0)A(t)\rangle]_{(\nu)}.
\end{align*}

In the other particular case $\langle A(t)B(0)\rangle=\langle B(t)A(0)\rangle$, which is the case for the quasi-particle and environment correlators we deal with (due to the electron/hole symmetry for quasiparticles, and to the gaussian character of phase fluctuations) on recovers immediately the same relation between positive- and negative- frequency Fourier transforms:
\begin{align*}
\mathcal{FT}[\langle A(t)B(0)\rangle]_{(\nu)}=& e^{-\beta h \nu}\mathcal{FT}[\langle B(0)A(t)\rangle]_{(\nu)}\\
=& e^{-\beta h \nu}\mathcal{FT}[\langle B(-t)A(0)\rangle]_{(\nu)}\\
=& e^{-\beta h \nu}\mathcal{FT}[\langle A(-t)B(0)\rangle]_{(\nu)}\\
=&e^{-\beta h \nu}\mathcal{FT}[\langle A(t)B(0)\rangle]_{(-\nu)}.
\end{align*}

\section{Details on sample}

The sample is the same as one of the samples used in \cite{AltimirasNoiseDCBPRL2014}. It consists in a quarter-wavelength resonator, which inner conductor consists in a series SQUID array. The Josephson inductance of the SQUIDs outranges the electromagnetic inductance by two orders of magnitude, bringing the characteristic impedance of the resonator in in the k$\Omega$ range. The resonator is terminated by a 270 k$\Omega$ Cu/AlOx/Cu tunnel junction.  In addition, a 30 $\times$ 50
$\times$ 0.3 $\mu$m $^3$ gold patch is inserted between the tunnel
junction and the SQUID array in order to evacuate the Joule power dissipated
at the tunnel junction via electron-phonon coupling. We briefly recall here details on the sample fabrication:

The 300 nm thick gold ground
plane of the resonator and thermalization pad were obtained by optical lithography, followed by evaporation 
and lift-off. 
SQUIDs
where fabricated following the process described in Ref.
\cite{PopReproducibleJunctionsArXiv2012}: the SQUIDs
(see the top inset) are obtained by double angle deposition of
($20/40\,\mathrm{nm}$) thin aluminum electrodes, with a $20'$
oxidation of the first electrode at $400\,\mathrm{mBar}$ of a ($85\%\,\mathrm{O}_2/15\%\,\mathrm{Ar}$) mixture. 
Before the evaporation, the substrate was cleaned by rinsing in ethanol and Reactive Ion Etching in an oxygen plasma \cite{Plasma}. The normal junction was
obtained using the same technique, with 30/60 nm thick copper electrodes and
an aluminum oxide tunnel barrier (5 nm thick aluminum oxidized for 15 minutes
at a 800 mBar (85\%O$_{2}$, 15\%Ar) mixture). 

\section{Details on the Josephson transmission line}

Our resonator consists in a $360\,\mathrm{\mu m}$ long
Josephson meta-material line containing 72 lithographically identical and
evenly spaced SQUIDs with a $5\,\mathrm{\mu m}$ period.   
The SQUIDs tunnel barriers have an area of
$0.5\,\mathrm{\mu m^2}$ each resulting in a room temperature
tunnel resistance $R_\mathrm{N}=720\,\Omega$. To assess that the SQUIDs in the array are identical, 
we have performed reproducibility tests,
yielding constant values of $R_\mathrm{N}$ (within  a few $\%$) over millimetric distances. Assuming a superconducting gap $\Delta=180\,\mathrm{\mu
  eV}$ and a 17\% increase of the tunnel resistance between room temperature and base temperature \cite{GloosGTTunnelAPL}, one obtains a zero flux critical current for the SQUIDs $I_C= 671$nA, 
  corresponding to $L_J(\phi=0)=0.49$ nH. 
 This corresponds to an effective lineic inductance   $\mathcal{L}\simeq\mathrm{100\,\mu H.m^{-1}}$ at zero
magnetic flux and frequency much lower than the Josephson plasma frequency of the junctions $\nu_\textrm{P}$\cite{RevModPhys.36.216}. Assuming  a capacitance for the junctions of the order of 80 fF/$\mu$m$^2$ yields $\nu_\textrm{P}\simeq$ 25 GHz.
Note that our simple fabrication mask produces 10 times bigger
Josephson junction in between adjacent SQUIDs, resulting in an additional $\sim \mathcal{L}\simeq\mathrm{10\,\mu H.m^{-1}}$ lineic  inductance. The  $\sim \mathcal {L}\simeq\mathrm{1\,\mu H.m^{-1}}$ electromagnetic inductance associated to our geometry is negligible.   With the designed lineic capacitance
$\mathcal{C}=75 \,\mathrm{pF.m^{-1}}$, the length of the resonator
sets the first resonance at $\nu_0\simeq 8\,\mathrm{GHz}$. The 12 fF shunting capacitance of the thermalization pad reduces these frequencies to $\nu_0 \simeq 6\,\mathrm{GHz}$.

\section{Details on setup and calibration}

\begin{figure}[h]
  \includegraphics[width=8cm]{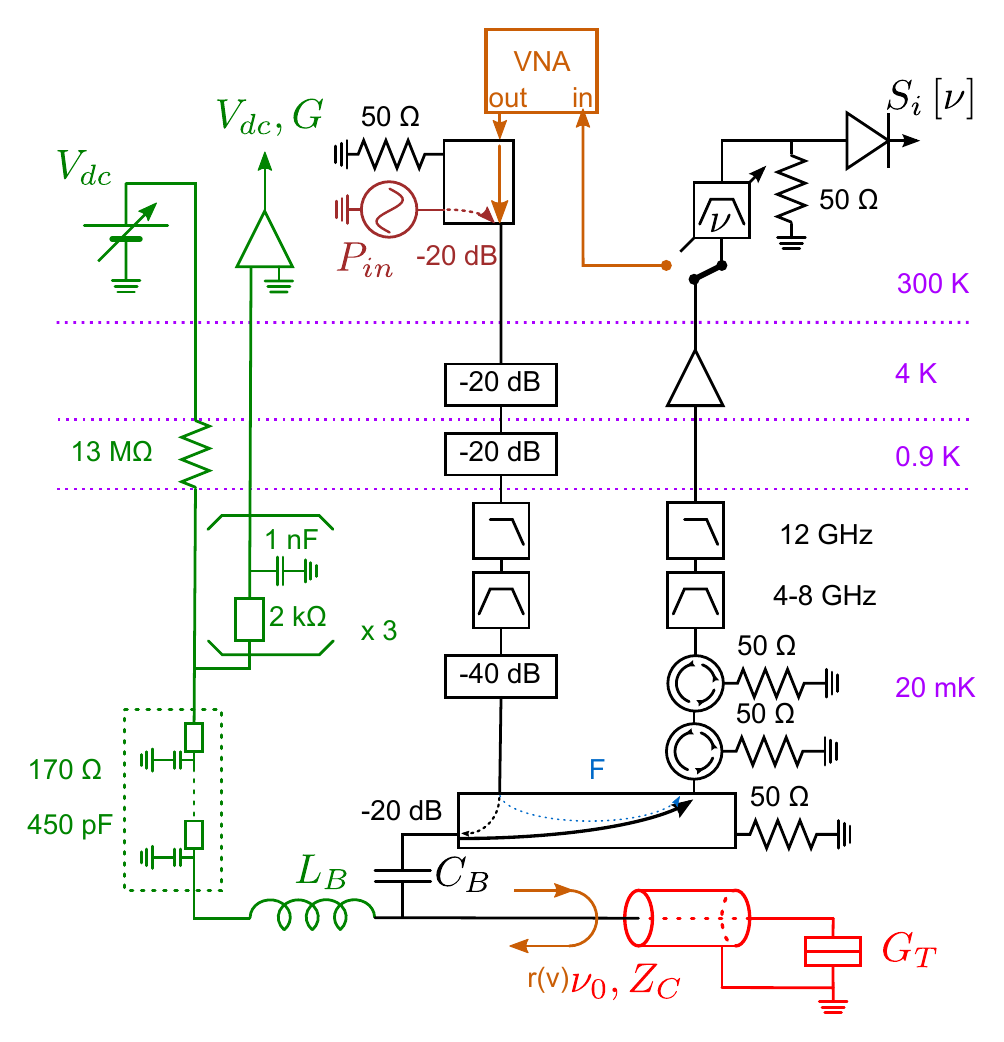}
  \caption{More detailed view of the experimental setup. Only circuit
  components inside the refrigerator are shown in full detail.}
	\label{FigureS2}
\end{figure}

We describe the calibration of the low frequency circuitry for voltage bias
and current measurement as well the microwave components used to define the
environment of the junction and to measure the emitted radiation.

\subsection{Low frequency circuit}

In addition to the components depicted in Fig.~1 of the main text, the low
frequency circuit includes a copper powder filter anchored on the mixing
chamber, as well as a distributed RC filter made with a resistive wire (50 cm
of IsaOhm 304 $\Omega$ m$^{-1}$) winded around a copper rod, and glued with
silver epoxy on a copper plate in good thermal contact with the mixing
chamber. Both are inserted between the 13 M$\Omega$ bias resistor and the bias
T and are represented by the 170$  \Omega$/450 pF RC filter on the biasing
line in Supplementary Material Fig. 1. The distributed RC filter has two
benefits on the effective electron temperature of our experiment: it provides
a high frequency filtering that reduces the polarization noise as well as
thermalization of the electrons. The copper powder filter is meant to absorb
parasitic microwave noise. The line allowing to measure the low frequency
response of the junction is filtered by a multipole RC low pass filter, made
with a succession of 2 k$\Omega$ Nickel-Chromium resistances and 1 nF
capacitances to ground. The NiCr resistances were checked in an independent
cool-down to change by less than 1\%, which allows to calibrate the 13
M$\Omega$ resistor in-situ, with a precision better than 1\%, which in turn
allows us to determine the dc voltage $V$ applied to the tunnel junction. The
validity of this calibration is confirmed by the quality of the comparison
between the observed steps in $\partial  S_{I} ( \nu ) / \partial  V$ and our
predictions.

\subsection{Microwave circuit and calibration}

The microwave chain comprises a bias Tee, two 4-8 GHz cryogenic circulators
anchored at the mixing chamber, as well as a 4-8 GHz bandpass filter and a 12
GHz low pass Gaussian absorptive filter (see Fig. \ref{FigureS2}).
These elements are anchored on the mixing chamber and are meant to protect the
sample from the back-action noise of the amplifier.

The quantitative
determination of the detection impedance relies on the detection of the power
emitted by the shot noise of the tunnel junction in the high bias regime. We
bias the junction at $\sim 1  \tmop{mV}$, where DCB corrections are
negligible, \ so that $S_{I} =2eI$ at frequencies $| \nu | \ll eV/h \simeq 0.5
\tmop{THz}$. In order to separate this noise from the noise floor of the
cryogenic amplifier, we then apply small variations of the bias voltage and
measure the corresponding changes in the measured microwave power with a
lock-in amplifier. The conversion of $S_{I}$ into emitted microwave power
depends on the environment impedance $Z ( \nu )$ seen by the tunneling
resistance $R_{T}$. First, only a fraction $R_{T}^{2} / |R_{T} +Z( \nu )|^{2}$
of the current noise is absorbed by the environment. The current noise in the
environment has then to be multiplied by $\tmop{Re}   [ Z( \nu ) ]$ to obtain
the microwave power emitted by the electronic shot noise:
\begin{equation}
  S_{P} ( \nu ) =2eV \frac{\tmop{Re}   [ Z( \nu ) ] G_{\Tau}^{-1}}{|Z( \nu
  )+G_{\Tau}^{-1} |^{2}} \simeq 2eV \frac{\tmop{Re}   [ Z( \nu ) ]
  G_{\Tau}^{-1}}{[ \tmop{Re}   [ Z( \nu ) ] +G_{\Tau}^{-1} ]^{2}} .
  \label{eq:shotnoise}
\end{equation}
The last approximation, $\tmop{Im}   [ Z( \nu ) ] \ll G_{\Tau}^{-1}$ is
satisfied with a precision better than 2\%. Finally, what is actually detected
at room temperature is the amplified microwave power:
\begin{equation}
  S^{\tmop{RT}}_{P} ( \nu ) =2eV G ( \nu ) \frac{\tmop{Re}   [ Z( \nu ) ]
  G_{\Tau}^{-1}}{[ \tmop{Re}   [ Z( \nu ) ] +G_{\Tau}^{-1} ]^{2}} .
  \label{eq:shotnoiseamplified}
\end{equation}
Supplementary Material Eq. \ref{eq:shotnoiseamplified} shows that the extracted
$\tmop{Re} [ Z( \nu ) ]$ depends on the gain of the microwave chain $G ( \nu
)$, which has to be determined in-situ and independently. To do so, we
inserted a 20 dB directional coupler between the sample and the bias Tee, and
injected through an independently calibrated injection line, comprising 70 dB
attenuation distributed between 4.2 K and the mixing chamber temperature (see
Supplementary Material Fig. 1). Both the attenuators and the directional
coupler were calibrated at 4.2 K. The gain of the microwave chain can be calibrated in situ, as explained below.

\subsection{Reflection measurement}

Due to the finite $\sim 15\,\mathrm{dB}$ directivity of the coupler, a coherent leak adds up to the microwave signal reflected by the sample, so that the total transmission coefficient can be expressed as $S_{21}(\nu_0)=G\left( \Gamma +F\right)$ where $\Gamma$ stands for the sample reflection coefficient at the input of the resonator, $F$ the coherent leak transmission, and $G$ stands for the total gain of the chain (including the attenuation of the various microwave components). We explain here how we measure independty $F$ and $G$.
We first apply a 200 $\mu$V dc bias to the sample, ensuring that the sample's admittance is given by $G_T$. 
By extrapolating the resonator's frequency with applied flux, we  set the resonator's frequency at $\nu_{\mathrm{detuned}}$ =3.7 GHz, so that the tunnel junctions impedance seen from the input of the resonator reads 

\begin{equation*}
Z_{\mathrm{detuned}}=
Z_0 \frac{1+ i G_T Z_0 \tan (\frac{\pi \nu_{\mathrm{detuned}}}{ 2 \nu_0})} {G_T Z_0 + i  \tan (\frac{\pi \nu_{\mathrm{detuned}}}{2 \nu_0})}, .
\end{equation*}

where $Z_0$ is the wave impedance of the SQUID's transmission line. The reflexion coeffiscient thus reads: 

\begin{widetext}
\begin{equation*}
\Gamma_{\mathrm{detuned}}=\frac{50 \Omega -Z_{\mathrm{detuned}}}{50 \Omega -Z_{\mathrm{detuned}}}=\frac{50 \Omega \left(G_T Z_0 + i  \tan (\frac{\pi \nu_{\mathrm{detuned}}}{2 \nu_0})\right)-Z_0 \left(1+ i G_T Z_0 \tan (\frac{\pi \nu_{\mathrm{detuned}}}{ 2 \nu_0})\right)}
{50 \Omega \left(G_T Z_0 + i  \tan (\frac{\pi \nu_{\mathrm{detuned}}}{2 \nu_0})\right)+Z_0 \left(1+ i G_T Z_0 \tan (\frac{\pi \nu_{\mathrm{detuned}}}{ 2 \nu_0})\right)}.
\end{equation*}
\end{widetext}

Note that since $G_T Z_0 \ll 1$, $\Gamma_{\mathrm{detuned}} \simeq -1$.
We then set the resonator's frequency back to $\nu_0$, while keeping the 200$\mu$V bias on the sample, so that the reflection coeffiscient reads $\Gamma_{\infty}= \displaystyle{\frac{50 \Omega-Z_0^2G_T}{50 \Omega+Z_0^2G_T}}$. From  $S_{21,\mathrm{detuned}}$ and $S_{21,\infty}$ we can deduce $G$ and $F$, allowing us to extract $\Gamma(V)=\displaystyle{\frac{50 \Omega-Z_0^2 Y(\nu_0, V)}{50 \Omega+Z_0^2Y(\nu_0, V)}}$, where $Y(\nu_0, V)$ stands for the complex admittance at the resonator's frequency of the junction biased at voltage $V$.

\subsection{Extracting the current noise}

We discuss here the possible consequences of the fact that the detection
impedance is not negligible compared to the tunneling resistance. More
specifically, we show that due to the variations of the tunneling resistance
with bias voltage, measuring $\partial  S_{P} ( \nu ) / \partial  V$ is not
rigourously equivalent to measuring $\partial  S_{I} / \partial  V$. However,
the error introduced by this approximation can be shown to be negligible.

Due to the non linearity of the tunnel transfer, the power emitted by the
junction biased at bias $V_{dc}$  reads
\begin{equation}
  S_{P} ( \nu ) = \tmop{Re}   [ Z( \nu ) ] \left| \frac{1} {1+ Y(\nu, V) Z(\nu)} \right|^{2} S_{I} ( V, \nu )
  \label{eq:sp} .
\end{equation}
Here $Y(\nu, V)$ is the differential admittance of the junction,
biased at voltage $V$, at the measurement frequency $\nu$. Supplementary
Material Eq. \ref{eq:sp} is valid as long as the ac current going through the
junction as a consequence of the shot noise is small enough for the response
of the junction $Y(\nu, V)$ to remain in the linear regime. In
that case, the modulation of the output voltage of the quadractic detector
that we measure is proportional to
\begin{equation}
\begin{split}
  \frac{\partial  S_{P} ( \nu )}{\partial  V_{dc}} = \tmop{Re}   [ Z( \nu ) ]
\left[ \left| \frac{1} {1+ Y(\nu, V) Z(\nu)} \right|^{2}  \frac{\partial S_{I} ( V, \nu )}{\partial  V} \right.\\ \left. +S_{I} ( V,
  \nu ) \frac{\partial}{\partial V}\left| \frac{1} {1+ Y(\nu, V) Z(\nu)} \right|^{2} \right] \label{eq:dspdv} .
	\end{split}
\end{equation}

From the measured variations of  $Y(\nu, V)$, we estimate that the associated
corrections are negligeable, so that detecting $\frac{\partial  S_{P} ( \nu
)}{\partial  V}$ gives direct access to $\frac{\partial S_{I} ( V, \nu
)}{\partial  V}$ within a precision better than 1\%.

\section{Photon population of the resonator induced by shot noise}

One can get a rough estimate of the photon population induced by shot noise. The photon emission rate density reads
\begin{align}
  \gamma(\nu)&=\frac{S_{P} ( \nu )}{h\nu} = \tmop{Re}   [ Z( \nu ) ] \left| \frac{1} {1+ Y(\nu, V) Z(\nu)} \right|^{2}\frac{S_{I} ( V, \nu )}{h\nu}
  \label{eq:sphot} .
\end{align}
 
As $Y(\nu, V) Z(\nu) \ll 1$, we get an estimate of the photon emission rate density by neglecting DCB effect on shot noise and a zero temperature:

\begin{equation}
 \gamma(\nu) \simeq 2 G_T \tmop{Re}   [ Z( \nu ) ] \frac{eV-h\nu}{h\nu}
\end{equation}

The average number of photons $\overline  n$  within the resonator can be estimated by intergrating the photon emission rate density, multiplied by the cavity lifetime $1/ (2\pi \Delta \nu)$, where $\Delta \nu$ is the FWHM of the resonator's impedance $ \tmop{Re}   [ Z( \nu ) ] $. For a representative bias voltage $eV=2 h\nu_0$, where $\nu_0$ is the resonant frequency:

\begin{equation}
 \overline  n \simeq G_T \tmop{Re}   [ Z( \nu_0 ) ] \frac{1}{\pi} \sim 0.02
\end{equation}

This parasitic population is small enough not to change the correlation functions of the environment detailed in \cite{AltimirasNoiseDCBPRL2014}.

\end{document}